\documentclass[a4paper,12pt]{article}
\usepackage{graphicx}
\usepackage{amsmath}
\usepackage{amssymb}

\begin{document}
\begin{titlepage}

\rightline{{\large \tt April 2002}}

\vskip 1.7 cm

\centerline{\Large \bf
The neutrino puzzle in the light of SNO
}

\vskip 1.4 cm

\centerline{\large R. Foot\footnote{foot@physics.unimelb.edu.au}  
and R. R. Volkas\footnote{r.volkas@physics.unimelb.edu.au}}
\vskip 0.7 cm\noindent

\centerline{{\large \it School of Physics}}

\centerline{{\large \it Research Centre for High Energy Physics}}

\centerline{{\large \it The University of Melbourne}}

\centerline{{\large \it Victoria
3010 Australia }}

\vskip 1.6cm

\centerline{\large Abstract} 
\vskip 0.7cm 
\noindent
SNO's neutral current measurement has added a new piece to the 
emerging neutrino physics puzzle. Putting together the 
presently available experimental information, 
an essentially unique picture emerges:
The solar neutrino anomaly is 
explained by $\nu_e \to \nu_\tau$ oscillations, the
atmospheric neutrino anomaly is explained by $\nu_\mu \to \nu_s$
and the LSND data are
explained by $\nu_e \to \nu_\mu$. This scheme
will be tested by future experiments: MiniBooNE will test
the oscillation explanation of the LSND anomaly, while the
long baseline experiments will discriminate between the
$\nu_{\mu} \to \nu_s$ and $\nu_{\mu} \to \nu_{\tau}$
possibilities for resolving the atmospheric anomaly
(confirming or disconfirming 
the Super-Kamiokande result that the latter is
favoured over the former).

\end{titlepage}

Over the last decade or so, important insight into the behaviour
of neutrinos has been gained from a number of important
experiments. We now have strong evidence for three neutrino
anomalies:

\begin{itemize}
\item
$\nu_\mu$ disappearance: Super-Kamiokande has established that the number
of up-going contained $\nu_\mu$ events is approximately half that of the
down-going events, while there is no discernible up-down asymmetry 
in the $\nu_e$-induced signal. These data 
can be explained by maximal
oscillation of $\nu_\mu$ into another species 
that cannot be $\nu_e$\cite{sk1}. Results from
Soudan II support this conclusion\cite{soudan}.

\item
SNO has now measured both the charged-current (CC) 
and neutral current (NC) solar neutrino rates, finding the flux ratio
$\phi^{NC}/\phi^{CC} = 2.9 \pm 0.4$\cite{sno}. 
These measurements are
consistent with other solar neutrino experiments\cite{othersolar}, 
and suggest
large mixing angle\footnote{Although the small-angle MSW solution is
disfavoured by the apparently undistorted nature of the $^8$B neutrino
spectrum, the overall $\chi^2_{min}$ of $99.9$ for $78$ degrees of freedom,
implies a C.L.\ of $4.8\%$ \cite{smirnov}, which 
is acceptable, so a small-angle
solution is still possible.}  
oscillations of $\nu_e$
into $\nu_{\mu}$ or $\nu_\tau$, with $\sin^2 2\theta \stackrel{>}{\sim}
0.7$ and a range of possible $\delta m^2$. (If all solar
neutrino results are taken at face value, then $\delta m^2 \sim 
10^{-4}\ eV$ is favoured. The SNO data on their own
allow a wider range.) 

\item
The LSND collaboration has obtained strong evidence 
for $\overline{\nu}_{\mu} \to \overline{\nu}_e$ oscillations via an
appearance experiment. They find $\delta m^2 \sim
1 \ eV^2$ and $\sin^2 2\theta \sim 10^{-2}$\cite{lsnd}.

\end{itemize}

If we assume that no light neutrinos exist beyond the known
$\nu_e, \nu_\mu$ and $\nu_\tau$ active flavours, then one is led to the
embarrassing conclusion that at least one piece of 
evidence must be discarded since
all three neutrino anomalies cannot be reconciled.
It is usually assumed that the LSND anomaly is due to
an unidentified error rather than oscillations,
with the solar and atmospheric neutrino anomalies then used as
evidence for a close-to bi-maximal mixing pattern 
amongst the three active neutrinos\cite{bim}.
Unfortunately, no compelling theoretical explanation for a
bi-maximal-like pattern has emerged.
On the other hand, there are simple schemes,
including mirror symmetry and pseudo-Dirac structure, that
can explain the existence of {\it two}-flavour maximal mixing.
Furthermore, the mirror symmetry idea actually predicts
the existence of light effectively sterile neutrinos,
so it is worth taking the possible existence 
of such additional species seriously.

It is quite possible that the three 
neutrino anomalies are all explained by oscillations.
If this is the case, then it is interesting that we
are led to an essentially unique picture:
LSND implies small angle $\nu_e \to \nu_\mu$ with 
$\delta m^2 \sim
1 \ eV^2$, SNO and the other solar neutrino experiments then 
suggest that $\nu_e$ oscillates
into $\nu_\tau$ with $\sin^2 2\theta \stackrel{>}{\sim} 0.7$,
for a range of $\delta m^2$. 
This leaves atmospheric $\nu_\mu$ disappearance
to be explained by approximately (or indeed exactly)
maximal $\nu_\mu \to \nu_s$ oscillations. 
We formalise these observations by proposing the following 
hypothesis: {\it The fundamental theory of neutrino mixing,
whatever it is, features (i) large (or even maximal)
$\nu_{\mu}$-$\nu_s$ mixing, (ii) small-angle active-active
mixing except for the $\nu_e$-$\nu_{\tau}$ channel which is
large.}

Interestingly, this scheme has {\it not} been considered 
previously in the literature, perhaps because the
Super-Kamiokande atmospheric data 
prefer the $\nu_\mu \to
\nu_\tau$ channel {\it in comparison with} 
the $\nu_\mu \to \nu_s$ possibility.
However, what really needs to be done to examine our hypothesis is
not to compare the $\chi^2_{min}$ of the $\nu_\mu \to \nu_s$ solution 
with the $\chi^2_{min}$ of the $\nu_\mu \to \nu_\tau$ solution, but rather,
to compute the allowed C.L.\ of the $\nu_{\mu} \to \nu_s$
hypothesis and check whether the goodness-of-fit
is reasonable. When this is done, the $\nu_\mu \to \nu_s$ solution 
{\it is} found to be reasonable, being only
mildly disfavoured, at the $1.5-3\sigma$ level, depending on how the data
are analysed\cite{sk2,foot}.
For example, a recent\cite{recent} Super-Kamiokande global fit for $\nu_\mu \to \nu_s$
oscillations has found a $\chi^2_{min}$ of $222.7$ for $190$ degrees of freedom.
This corresponds to an allowed C.L.\ of 5.5\%, that is, it is only mildly
disfavoured at about the $2\sigma$ level.

There are two possibilities for the mass hierarchy:
either $\nu_e, \nu_\tau$ are 
\linebreak
$\sqrt{\delta m^2_{LSND}} \sim 1 \ eV$
heavier than $\nu_\mu, \nu_s$ or vice-versa.
If the $\nu_e, \nu_\tau$ pair is heavier, 
then it might put the $\nu_e$ mass, assuming it is Majorana,
in the range accessible to neutrinoless double beta decay 
experiments.\footnote{However, if 
the large-angle between $\nu_e$ and $\nu_{\tau}$ is due to
an approximate
pseudo-Dirac character, then neutrinoless double
beta decay may occur too slowly to be seen.}

In the nearer future, the two most important tests will come
from the MiniBooNE experiment and the long baseline
experiments MINOS and CNGS. MiniBooNE has been designed
to check the oscillation explanation of the LSND result, 
and with it the indirect evidence for a light sterile flavour.
Through NC or $\tau$-appearance measurements, the up-coming 
long baseline experiments hope to
discriminate between the $\nu_{\mu} \to \nu_{\tau}$ and
$\nu_{\mu} \to \nu_s$ channels used to resolve the
atmospheric anomaly. This will provide a terrestrial check
of the Super-Kamiokande result that the $\nu_{\tau}$
possibility is favoured.

On the theoretical front,
note that this four-neutrino phenomenological scheme can be
embedded into interesting six-neutrino theories, such as provided
by the mirror matter model\cite{ms}. 
Recall that this theory predicts that each ordinary neutrino, 
$\nu_{\alpha}$ where $\alpha = e,\mu,\tau$,
is maximally mixed with an effectively sterile mirror
partner $\nu'_{\alpha}$. The interfamily mixing pattern,
the absolute mass scales,
and the $\nu_{\alpha}-\nu'_{\alpha}$ mass splittings are
all free parameters.
The ordinary-mirror neutrino sector will reduce to the above
four-neutrino scenario if the oscillation
lengths of the $\nu_e \to \nu'_e$ and $\nu_\tau \to \nu'_\tau$ modes 
are greater than the Earth-Sun distance at solar neutrino energies. 
(It might be possible to relax this assumption, but further work
needs to be done.) The importance of this possible theoretical basis 
for the scheme is twofold. First, the presence of a light effectively 
sterile neutrino (the $\nu'_{\mu}$) is explained, because mirror
neutrinos are light for exactly the same reason that ordinary 
neutrinos are light (whatever that reason might be). Second,
the atmospheric mixing is predicted to be exactly pairwise
maximal, unlike the $\nu_e - \nu_{\tau}$ mixing angle which
could be just accidentally large. These features can also
be more stringently tested in the future.
For example,
the Japan Hadron Facility to Super-Kamiokande superbeam
proposal cites an experimental precision of about 0.01
on the atmospheric $\sin^2 2\theta$ as achievable \cite{jhf}.
KAMLAND may soon provide more information on the solar mixing angle.

In summary, the most important features of the three
neutrinos anomalies -- solar, atmospheric and LSND --
can be explained through large angle $\nu_e \to \nu_{\tau}$,
exactly or approximately maximal $\nu_{\mu} \to \nu_s$,
and small angle $\nu_e \to \nu_{\mu}$ oscillations, respectively.
The new information from SNO, when combined with the
LSND result, suggests that the solar channel is 
predominantly $\nu_{\tau}$. Important aspects of
this scenario will be tested soon by MiniBooNE
and the long baseline experiments. When embedded within
the mirror matter model, an important difference
between the large solar and the large atmospheric
mixing angles emerges: the former is large for
possibly accidental reasons, while the latter is
rigorously maximal for theoretical reasons. Future
precision measurements of $\nu_{\mu}$ disappearance
in superbeam experiments will test that possibility
further.



\newpage

\begin{thebibliography}{999}
\bibitem{sk1}
Super-Kamiokande Collaboration,
Y. Fukuda {\it et al.}, Phys.
Rev. Lett. 82, 1562 (1998); Phys. Lett. B436, 33 (1998);
Phys. Lett. B433, 9 (1998).

\bibitem{soudan}
Soudan 2 Collaboration,
W. W. Allison {\it et al.}, 
Phys. Lett. B449, 137 (1999).

\bibitem{sno}
SNO Collaboration,
Q. R. Ahmad {\it et al.}, nucl-ex/0204008; nucl-ex/0204009;
Phys. Rev. Lett. 87, 071301 (2001).

\bibitem{othersolar}
Homestake Collaboration, B. T. Cleveland {\it et al.}, Astrophys. J.
496, 505 (1998); Kamiokande Collaboration, Y. Fukuda {\it et al.},
Phys. Rev. Lett. 77, 1683 (1996); Super-Kamiokande
Collaboration, Phys. Rev. Lett. 86, 5651 (2001);
Sage Collaboration, J. N.
Abdurashitov,
{\it et al.}, Phys. Rev. Lett. 83, 4686 (1999);
Gallex Collaboration, W. Hampel {\it et al.}, Phys. Lett. B447,
127 (1999); GNO Collaboration, M. Altann {\it et al.}, Phys. Lett.
B490, 16 (2000).

\bibitem{smirnov}
See Table 1 of
P. C. de Holanda and A. Yu.\ Smirnov, hep-ph/0205241.

\bibitem{lsnd}
LSND Collaboration,
C. Athanassapoulos {\it et al.}, Phys. Rev. Lett. 81, 1774 (1998);
Phys. Rev. C58, 2489 (1998).

\bibitem{bim}
V. Barger {\it et al.}, Phys. Lett. B437, 107 (1998); A. Baltz,
A. S. Goldhaber and M. Goldhaber, Phys. Rev. Lett. 
81, 5730 (1998); F. Vissani, hep-ph/9708483;
D. V. Ahluwalia, Mod. Phys. Lett. A13, 2249 (1998).


\bibitem{sk2}
Super-Kamiokande Collaboration,
S. Fukuda {\it et al.},
Phys. Rev. Lett. 85, 3999 (2000).

\bibitem{foot}
R. Foot, Phys. Lett. B496, 169 (2000).

\bibitem{recent}
M. Shiozawa for the Super-Kamiokande Collaboration, talk at
the Neutrino2002 conference, Munich, May 2002.

\bibitem{ms}
R. Foot, H. Lew and R. R. Volkas, Mod. Phys. Lett. A7, 2567 (1992);
R. Foot, Mod. Phys. Lett. A9, 169 (1994);
R. Foot and R. R. Volkas, Phys. Rev. D52, 6595 (1995).

\bibitem{jhf}
Y. Itow {\it et al.}, hep-ex/0106019.

\end{thebibliography}
\end{document}